\documentclass[12pt]{article}
\usepackage{epsf}
\newcommand{\be}{\begin{equation}}
\newcommand{\ee}{\end{equation}}
\newcommand{\bea}{\begin{eqnarray}}
\newcommand{\eea}{\end{eqnarray}}
\newcommand{\brr}{\begin{array}}
\newcommand{\err}{\end{array}}
\newcommand{\bc}{\begin{center}}
\newcommand{\ec}{\end{center}}
\newcommand{\nn}{\nonumber}

\newcommand{\ep}{\varepsilon}

\newcommand{\epse}{\varepsilon^{\prime}/\varepsilon}
\renewcommand{\Im}{\mbox{Im}}
\renewcommand{\Re}{\mbox{Re}}
\newcommand{\as}{\alpha_s}
\newcommand{\cO}{{\cal O}}

\newcommand{\lsim}{\stackrel{<}{_\sim}}

\newcommand{\PL}[3]{{\it Phys.\ Lett.\ }         {\bf #1}, {#3} {(19#2)}}

\newcommand{\PRL}[3]{{\it Phys.\ Rev.\ Lett.\ }  {\bf #1}, {#3} {(19#2)}}
\newcommand{\PR}[3]{{\it Phys.\ Rev.\ }          {\bf #1}, {#3} {(19#2)}}
\newcommand{\NP}[3]{{\it Nucl.\ Phys.\ }         {\bf #1}, {#3} {(19#2)}}

\newcommand{\JHEP}[3]{{\it JHEP\ }               {\bf #1}, {#3} {(19#2)}}
\newcommand{\IJMP}[3]{{\it Int. J. Mod. Phys.\ } {\bf #1}, {#3} {(19#2)}}

\newcommand{\ibid}[3]{{\it ibid.\ }              {\bf #1}, {#3} {(19#2)}}
\begin{document}
\pagestyle{empty} 
\begin{flushright}
INFNNA-99/39 \\
LPT ORSAY 99/99
\end{flushright}
\vskip 1cm
\centerline{\Large{\bf{Direct CP violation in $K \to 3\pi$ decays}}}
\centerline{\Large{\bf{induced by SUSY chromomagnetic penguins$^*$}}}
\vskip 1cm
\centerline{\bf{G.~D'Ambrosio$^a$, G. Isidori$^{b}$ and G. Martinelli$^{c,d}$}}
\vskip 0.3cm
{\small
\centerline{$^a$ INFN, Sezione di Napoli and Dipartimento di Scienze 
 Fisiche, }
\centerline{Universit\`a di Napoli, I-80125 Napoli, Italy}
\centerline{$^b$ INFN, Laboratori Nazionali di Frascati, 
 P.O. Box 13, I-00044 Frascati, Italy}
\centerline{$^c$ LAL and Laboratoire de Physique Th\'eorique,}
\centerline{Laboratoire associ\'e Centre National de la Recherche
Scientifique,}
\centerline{
Universit\'e de Paris XI, B\^at 211, 91405 Orsay Cedex, France}
}
\begin{abstract}
An analysis of the CP violating asymmetry in $K^\pm \to (3\pi)^\pm$ decays
in the Standard Model and, by means of the mass insertion approximation, 
in a wide class of possible supersymmetric extensions, is presented.
We find that the natural order of magnitude for this asymmetry is 
$\cO(10^{-5})$ in both cases.  Within supersymmetric models  
effects as large as $\cO(10^{-4})$ are
possible, but only in a restricted range of the relevant parameters.
\end{abstract}
\vfill
\begin{flushleft}
November 1999
\end{flushleft}
\vfill
\noindent
{\footnotesize $^*$ Work supported in part by TMR, EC--Contract No. 
ERBFM RX--CT980169 (EURODA$\Phi$NE).} \\
{\footnotesize $^d$ On leave of absence from Dipartimento di Fisica, 
 Universit\`a di Roma ``La Sapienza'' and INFN, Sezione di Roma, 
 P.le A. Moro 2, I-00185 Rome, Italy. }
\newpage
\pagestyle{empty}\clearpage
\setcounter{page}{1}
\pagestyle{plain}
\newpage
\pagestyle{plain} \setcounter{page}{1}
\paragraph{1.}
The origin of CP violation is one of the fundamental 
questions of particle physics
and cosmology which  remains an open problem to date. 
The recent measurements of $\epse$ \cite{epsp99} (see also \cite{NA31})
represent an important step forward in our understanding
of this phenomenon, since they have ruled out superweak scenarios.
Nonetheless we are still far from a quantitative description
of the  dynamics which generate the  amount
of CP violation observed in  hadronic processes.
Indeed, even within the Standard Model (SM)
it is very hard to predict the value of
$\epse$ in terms of the parameters of the 
Cabibbo-Kobayashi-Maskawa (CKM) matrix
(see e.g. Refs.~\cite{epspth} and references therein). 
Given the large theoretical uncertainties 
affecting the calculation of this quantity, it is
very useful to collect additional experimental information 
about CP violation in $|\Delta S|=1$ transitions.
In this respect charge asymmetries in non-leptonic decays,
as the difference in  $K^+\to (3\pi)^+$ and 
$K^-\to (3\pi)^-$ Dalitz plot distributions \cite{Avilez},
represent an interesting class of observables, since they are 
straight direct CP-violating effects
 free from $|\Delta S|=2$ contaminations.
Moreover,  contrary to $\epse$, these asymmetries  stem from 
the interference of two $\Delta I=1/2$
amplitudes  and do not necessarily suffer  the suppression  
of $\Delta I=3/2$ transitions.
In spite of these advantages, however, within the SM such observables 
are expected to be very small, of $\cO(10^{-5})$,
due to the constraints from $\epse$ and the smallness 
of final-state interactions \cite{K3p1,K3p2}.
A natural question is whether extensions of the SM could enhance 
these CP-violating asymmetries at such a level that they  
could be recognized as a clear signal of new physics.
\par 
Good candidates to provide new large CP violating effects 
are the supersymmetric extensions of the SM with generic 
flavour couplings and minimal particle content. In this framework,
among the possible contributions which may be envisaged,
it has been recently recognized the importance of the 
chromomagnetic operator (CMO).
Its CP-odd contribution can become large in the
presence of misalignment between quark and squark mass matrices,
 and, without conflict with the experimental
determination of the $K^0$--$\bar K^0$ mixing amplitude,
it can account for the largest part of the measured $\epse$
\cite{MM}--\cite{BCIRS}.
Actually a non-standard CMO is the only
possibility, within this framework, to considerably affect the 
CP-violating part of $\Delta I=1/2$ amplitudes 
without serious fine-tuning problems in $|\Delta S|=2$ 
processes \cite{BCIRS}.
\par 
In this paper we investigate the possibility of using the CMO to 
enhance CP violating effects in $K \to 3 \pi$ decays. 
We work under the assumption that
the  Wilson coefficient of this operator is mostly due to 
left-right mixing among
down-like squarks. This implies that 
a large chromomagnetic term is necessarily accompanied 
by sizeable corrections to   $\ep$ and  $K_L \to \pi^0 e^+ e^-$.
We thus perform a combined analysis of these processes 
together with $\epse$ and $K \to 3 \pi$ decays. 
We have not considered other possible supersymmetric sources of CP violation
in $K \to 3 \pi$ decays, such as left-left or right-right squark mixing, since 
only the left-right ones trigger the enhancement of the CMO which is the most
promising candidate to give an observable effect.
\par 
Our main conclusions are  that, even within supersymmetry,
effects at the level of $\cO(10^{-4})$ may  be observed 
only under special circumstances,
among which  large cancellations of  different contributions  in $\epse$.
Otherwise, similar to the SM,  the natural order of magnitude of the charge
asymmetries will remain of $\cO(10^{-5})$.  
As we shall discuss, this conclusion is rather general
and applies also to other direct CP-violating observables in 
non-leptonic processes.
\par 
The paper is organized as follows. We first derive the main 
formulae needed to evaluate the CMO contribution to CP-odd 
asymmetries in $K^{\pm} \to (3\pi)^\pm$ decays.
Then we discuss the role of down-type left-right mass 
insertions, including contributions from the CMO, in the 
$K^0$--$\bar K^0$ mixing amplitude. Finally a combined analysis 
of   CP-violating effects in $K^{\pm} \to (3\pi)^\pm$ decays,
taking into account the constraints imposed by
$\ep$, $\epse$ and  $K_L \to \pi^0 e^+ e^-$, is presented.
The results are summarized in the conclusions.

\paragraph{2.} We start by analyzing the charge asymmetries 
in $K^{\pm} \to \pi^\pm\pi^\pm\pi^\mp$ decays.
As discussed in Ref.~\cite{K3p1,K3p2}, the most 
interesting CP-violating observable is the asymmetry
in the Dalitz plot slopes $g_\pm$.
Neglecting the suppressed $\Delta I=3/2$ contributions,
this can be written as 
\be 
\frac{g_+-g_-}{g_+ + g_-} = \left[ \frac{\Im b}{\Re b} -
\frac{\Im a}{\Re a} \right] \sin (\alpha_0 - \beta_0) \, , 
\label{eq:dg} 
\ee 
where the weak amplitudes $a$ and $b$ are defined by the 
momentum expansion of $A(K^{+} \to \pi^{+}\pi^+\pi^-)$
around the center of the Dalitz Plot,
\bea
A(K^{+} \to \pi^{+}\pi^+\pi^-) &=& a e^{i \alpha_0} + b e^{i \beta_0 } Y
+ \cO(Y^2)\, , \label{eq:Ak3p} \\ 
Y &=& \frac{3(p_K-p_{\pi^-})^2 - M_K^2 -3M_\pi^2}{M_\pi^2}\, , 
\eea
and $\alpha_0,~\beta_0$ 
are the small rescattering phases, known from 
chiral perturbation theory (ChPT) \cite{DIPP},  
evaluated at $Y=0$. In the limit where we neglect 
$\Delta I=3/2$ contributions, the slope asymmetries of
$K^{\pm} \to \pi^\pm\pi^\pm\pi^\mp$  and 
$K^{\pm} \to \pi^0\pi^0\pi^\pm$ modes are identical.

Since $a$ and $b$ are $\Delta I=1/2$ amplitudes, on general grounds one 
expects $\Im a/\Re a$ and  $\Im b/\Re b$ to be both of the same order as
the weak phase of $A_0=A(K\to(2\pi)_{I=0})$, namely
\be  
\frac{\Im a}{\Re a} \sim \frac{\Im b}{\Re b} \sim 
\frac{\Im A_0}{\Re A_0} \sim 
\Re\left(\frac{\varepsilon^\prime}{\varepsilon}\right)
\vert   \varepsilon \vert  \frac{\Re A_0}{\Re A_2}
\sim 10^{-4} \, .
\ee
Given that $\sin(\alpha_0 - \beta_0) \lsim 0.1$~\cite{DIPP}, 
this sets the ``natural" order of magnitude for the asymmetry to $10^{-5}$
\cite{K3p1}.
Actually, within the  SM,  the situation is even worse:
neglecting the CMO, which in this case has a very small coefficient,
the asymmetry vanishes  at the lowest order in the
chiral expansion. This happens because at this order there is only one 
octet operator which generate the same weak phase to all
the $\Delta I=1/2$ amplitudes \cite{K3p2}.   
Clearly the situation may improve if the  contribution of  
the CMO is enhanced by supersymmetric effects, which we now discuss. 
\par 
Let us start from the relevant piece of the effective Hamiltonian.
This can be written as \cite{BCIRS}
\be
{\cal H}_{\rm mag} = C^+_g Q^+_g + C^-_g Q^-_g + {\rm h.c.}~,
\label{eq:heff}
\ee
where
\be
Q^\pm_g = \frac{g}{16 \pi^2}
        \left( {\bar s}_L \sigma^{\mu \nu} t^a G^a_{\mu\nu} d_R \pm
               {\bar s}_R \sigma^{\mu \nu} t^a G^a_{\mu\nu} d_L \right)
\ee
and the dominant contribution to 
Wilson coefficients, generated by gluino exchange diagrams, 
is given by \cite{BCIRS,GGMS}
\be
C^\pm_g(m_{\tilde g}) = \frac{\pi \alpha_s(m_{\tilde g})}{m_{\tilde g}}
 \left[ \left(\delta^{D}_{LR}\right)_{21} \pm
   \left(\delta^{D}_{LR}\right)^*_{12} \right] G_0(x_{g q})\, .
\ee
Here $(\delta^D_{LR})_{ij}=(M^2_D)_{i_L j_R}/m^2_{\tilde q}$ 
denote the off-diagonal entries of the (down-type) squark mass 
matrix in the super-CKM basis \cite{HKR} and
$x_{g q}=m^2_{\tilde g}/m^2_{\tilde q}$ the ratio of 
gluino and (average) squark mass squared.
The explicit expression of  $G_0(x)$ can be found in 
\cite{BCIRS}.

\par
%
The realization of $Q^\pm_g$ in terms of meson fields,
to the lowest order in $1/N_c$ and in the derivative
expansion, can be written as
\be 
Q^\pm_g =  \frac{11}{256 \pi^2} \frac{f_\pi^2 M_K^2}{m_s+m_d} \left[
U D_\mu U^\dagger D^\mu U \pm  D_\mu U^\dagger D^\mu U U^\dagger
 \right]_{23} \, ,   \label{eq:CMOChPT} 
\ee
where $U =  \exp(i 2 \phi /f_\pi)$,  $f_\pi=132$ MeV,
$\phi$ is the octet field of pseudoscalar mesons,
and the overall coupling has been fixed by 
the chiral quark model estimate of Ref.~\cite{BEF}.
Using Eq.~(\ref{eq:CMOChPT}) we can derive the 
following matrix elements
\bea 
\langle  \pi^0(p) \vert Q^+_g \vert  K^0 (p) \rangle &= &
- \frac{11 B_{g1}}{32 \sqrt{2} \pi^2} \frac{M_K^2 p^2}{m_s+m_d} \, , 
  \label{eq:Bg1}   \\
i \langle \pi^+ \pi^- \vert Q^-_g \vert  K^0\rangle & = & 
- \frac{11 B_{g2}}{32  \pi^2} \frac{M_K^2 M_\pi^2}{f_\pi(m_s+m_d)} \, ,\\
  \langle \pi^+\pi^+ \pi^- \vert Q^+_g \vert  K^+\rangle & = &
- \frac{11 B_{g3}}{16 \pi^2} \frac{M_K^2 M_\pi^2}{f_\pi^2(m_s+m_d)} \,
. \label{eq:Bg3}
\eea
The $B$-factors, $B_{gi}$, have been introduced to 
parametrize our ignorance of the precise overall 
coefficient in Eq.~(\ref{eq:CMOChPT}) and of 
possible higher-order terms. For practical purposes,
in the following we shall 
also set $m_s+m_d=110$~MeV in Eqs.~(\ref{eq:Bg1})--(\ref{eq:Bg3}), 
encoding in the $B_{gi}$ the  remaining uncertainty on the true value of 
the quark masses.\footnote{~For an
extensive discussion about the possible chiral realizations of the CMO see 
Ref.~\protect\cite{HeV}. The factor 
$B_{g3}$ could in principle be  a function of $Y$, but for simplicity
in the following we will ignore this possibility;  $B_{g2}$ 
coincides with the $B_G$ of \protect{\cite{BCIRS}} for $m_s+m_d=110$~MeV.}
\par For the $K \to 3 \pi$ amplitudes defined in (\ref{eq:Ak3p})
we obtain
\be 
\left| \frac{\Im a}{\Re a} \right| =  \frac{3}{2 M_K^2 G_8} \times 
\frac{11 }{16 \pi^2} 
\frac{M_\pi^2 M_K^2}{f_\pi^2(m_s+m_d)}~\left| B_{g3} 
 \Im C^+_g\right| \, , \qquad
\quad \frac{\Im b}{\Re b} =  0\, ,
\label{eq:questa}\ee
where we have used the lowest-order chiral relation between 
$\Re a$  and $\Re A_0$, expressing the latter in 
terms of the standard coupling $G_8= 9.1 \times 10^{-6}$ GeV$^{-2}$
(i.e. $\Re a=  2 M_K^2 G_8/3$).
In view of the numerical analysis, it is convenient to introduce the following
simple expression, which can be readily derived from Eq.~(\ref{eq:questa})
and ${\cal H}_{\rm mag}$ in  (\ref{eq:heff})
\be 
\left| \frac{g_+-g_-}{g_+ + g_-} \right| \simeq 1.97 \times   \left[ 
\frac{\eta~ \as(m_{\tilde g})}{\as(
500 {\rm GeV})} \frac{500 {\rm GeV}}{m_{\tilde g}}
\frac{G_0(x_{gq})}{G_0(1)} \right]~ \left| 
 B_{g3} \Im \delta_+ \right| \label{eq:dgsg} \, ,
\ee
where we have defined
 $\delta_\pm = (\delta_{LR}^D)_{21} \pm (\delta_{LR}^D)^*_{12} = 
(\delta_{LR}^D)_{21} \pm (\delta_{RL}^D)_{21}$.
We found very useful to introduce $\delta_\pm$ since these are the  natural 
couplings appearing at first order in any  parity conserving ($+$) or
parity violating ($-$) observable.
In the evaluation of the numerical coefficient above we have used 
$\as(500 {\rm GeV})=0.096$ and, as anticipated, $m_s+m_d=110$ MeV. 
The parameter $\eta\simeq 0.9$ \cite{BCIRS} is the correcting factor 
due to the running of the Wilson
coefficient from $m_{\tilde g}$ to the operator renormalization scale.

\paragraph{3.} 
An important constraint on the couplings $\delta_\pm$ comes from 
$K^0$--$\bar  K^0$ mixing. Besides the usual gluino-box amplitudes widely
discussed in the literature (see e.g. 
Refs.~\cite{GGMS,susyboxes,bagger} and references therein) 
further contributions arise from the single and
double insertion of the CMO. 
Schematically we can write
\be 
{\cal A}^{\rm susy}(K^0 \to \bar K^0) = {\cal A}_{\rm boxes}
+{\cal A}_{\rm 1-mag} +{\cal A}_{\rm 2-mag}\, . 
\ee
${\cal A}_{\rm boxes}$, which is dominated by short-distance contributions,
can conveniently be written in the  form
\bea   
{\cal A}_{\rm boxes} &=& 
 \frac{\alpha_S^2}{m_{\tilde g}^2}  \frac{1}{432}
 \left(\frac{M_K}{m_s+m_d}\right)^2 M_K^2 f_K^2 \left[
  x_{gq}^2 f_6(x_{gq}) \left( 85\eta_2 B_2 +3
  \eta_3 B_3 \right)(\delta_+^2 +\delta_-^2) \right. \nn \\ 
  &&\qquad\qquad\qquad + \left. x_{gq} \tilde f_6(x_{gq}) \left(
 33 \eta_4 B_4 + 15  \eta_5 B_5\right)
  (\delta_+^2 - \delta_-^2) \right]~,
\eea
where the definitions of $f_6(x)$, $\tilde f_6(x)$ and of the $B_i$
parameters have been taken from Ref.~\cite{susyboxes}.
In the numerical analysis we will use the simplified expression
\be  
{\cal A}_{\rm boxes} = 2.9 \times 10^{-11} {\rm GeV}^2 \times \left[ 
\frac{\as(m_{\tilde g})}{\as(500 {\rm GeV})} 
\frac{500 {\rm GeV}}{m_{\tilde g}}\right]^2 \left( B_+
\, \delta^2_+\, +\, B_- \, \delta^2_- \right) 
 \label{eq:nboxes} \, , 
\ee 
where $B_\pm$ are coefficients of $\cO(1)$, which may vary by a factor
of $2 \div 3$  depending on the values of the $B$ parameters of 
the $\Delta S=2$ operators, the precise value of $x_{gq}$, 
the perturbative QCD corrections, etc.~\cite{susyboxes}.
\par 
The single and double insertions of the CMO  are expected to be dominated 
by long distance contributions. For illustration we give here the expression
of  the single $\pi^0$ contribution, already considered in Ref.~\cite{mura2}
\bea 
{\cal A}^{\pi^0}_{\rm 1-mag} &=&   2 \,\langle \bar K^0 \vert 
{\cal H}^{\Delta S=1}_{\rm SM} \vert \pi^0\rangle
\frac{1}{M_K^2-M_\pi^2}  \langle  \pi^0 \vert {\cal H}_{\rm{mag}}
\vert K^0\rangle
\nn \\ &=&  \frac{1}{M_K^2-M_\pi^2} \left( 
 \frac{11 B_{g1} }{32  \pi} \frac{M_K^6}{m_s+m_d} G_8 f_\pi^2 \right)  
 \frac{\alpha_S}{m_{\tilde g}}  \eta G_0(x_{gq})\delta_+ \, . 
\label{eq:DS21mag} 
\eea
One can then generalize the above expression to include 
the contribution from other
one-meson states, such as the $\eta$ and $\eta^\prime$. 
In the following we will use the simplified expression
\be  
{\cal A}_{\rm 1-mag} =  4.8 \times 10^{-13} {\rm GeV}^2 \times \left[
\frac{\eta~\as(m_{\tilde g})}{\as(
500 {\rm GeV})} \frac{500 {\rm GeV}}{m_{\tilde g}} 
\frac{G_0(x_{gq})}{G_0(1)}\right]
\kappa_1 \, \delta_+ \label{eq:n1mag} \, ,
\ee
where the numerical coefficient has been computed from 
Eq.~(\ref{eq:DS21mag}) and we have absorbed  the hadronic
uncertainties, namely $B_{g1}$ and contributions from 
intermediate states other than the $\pi^0$, in the factor $\kappa_1$.
Using  the Gell-Mann-Okubo mass formula the $\pi^0$ and $\eta$ contributions
would cancel, thus one may argue that $\kappa_1$ should be substantially
smaller than one. We know, however, that a similar argument fails for
$K_L \to \gamma \gamma$, where the effective coupling, corresponding
to our $\kappa_1$, is of ${\cal O}(1)$.  
Note that intermediate states with parity opposite to the one
pseudoscalar-meson state
may give contributions proportional to $\delta_-$. We have neglected these
effects in our analysis.
\par 
Similarly,  in the case of the double insertion one gets 
\be {\cal A}^{\pi^0}_{\rm 2-mag} =   \langle \bar K^0 \vert 
{\cal H}_{\rm mag} \vert \pi^0\rangle
\frac{1}{M_K^2-M_\pi^2}  \langle  \pi^0 \vert {\cal H}_{\rm mag} \vert K^0\rangle
\, ,\ee
that proceeding as before  leads to
\be  {\cal A}_{\rm 2-mag} = 1.9 \times 10^{-11} {\rm GeV}^2 \times 
\left[\frac{\eta~\as(m_{\tilde g})}{\as(
500 {\rm GeV})} \frac{500 {\rm GeV}}{m_{\tilde g}} 
\frac{G_0(x_{gq})}{G_0(1)}\right]^2
\kappa_2\, \delta^2_+ \label{eq:n2mag} \, . 
\ee

\paragraph{4.} For the other two quantities 
which are used in our analysis,
namely $\epse$ and the ${\rm BR}(K_L \to \pi^0 e^+ e^-)$, 
rather than giving the explicit analytic
expressions, for which we refer the reader to Ref.~\cite{BCIRS},
we only list here two convenient expressions:
\be  
\Re\left(\frac{\varepsilon^\prime}{\varepsilon} \right)_{\rm mag}
=  92.6 \times  \left[ 
\frac{\eta~\as(m_{\tilde g})}{\as(
500 {\rm GeV})} \frac{500 {\rm GeV}}{m_{\tilde g}} 
\frac{G_0(x_{gq})}{G_0(1)}\right]
B_{g2}  \, \Im\delta_- \label{eq:nepse1mag} \, , \ee
\be 
{\rm BR}(K_L \to \pi^0 e^+ e^-)_{\rm mag} =
6.1 \times 10^{-4} 
\left(\frac{\tilde y_\gamma(m_{\tilde g}, x_{gq}) G_0(x_{gq})}
{\tilde y_\gamma(500 {\rm GeV}, 1) G_0(1)}\right)^2 
B_T^2 (\Im \delta_+)^2 \, ,  \label{eq:nschifo}
\ee
where  the definitions of $B_T$ and $\tilde y_\gamma$ can be found 
in Ref.~\cite{BCIRS}, and in $K_L \to \pi^0 e^+ e^-$ we have 
neglected the interference with the SM contribution.
\par 
Besides the numerical expressions given above, for our study we also use
the following experimental inputs:
\bea 
&& | \ep |~=~ ( 2.28 \pm 0.02 ) \times 10^{-3}~, \qquad 
\Re\left(\frac{\varepsilon^\prime}{\varepsilon} \right)~=~ (21.2
\pm 4.6) \times 10^{-4}~, \qquad \\
&& {\rm BR}(K_L \to \pi^0 e^+ e^-)~<~ 5.6 \times 10^{-10}
\label{eq:exps} \,  ~\protect\cite{KTeVrari}~.
\eea
\par 
We are now ready to discuss the bounds in the 
$\Im \delta_+$--$\Im \delta_-$ plane
imposed by the experimental measurements and the theoretical
expressions given in Eqs.~(\ref{eq:nboxes}), (\ref{eq:n1mag}), 
(\ref{eq:n2mag})--(\ref{eq:nschifo}).
As usual, we impose the constraints by requiring that 
all CP violating observables are saturated by
the supersymmetric effects considered above, i.e. neglecting 
the SM contributions.
By setting the values of all hadronic parameters (namely the $B_i$'s and the
$\kappa_i$'s) to one, at our reference values of gluino and squark masses
($m_{\tilde g}=500$ GeV and $x_{gq}=1$), 
we find $\Im \delta_- \sim \Im \delta_+ 
\sim 10^{-5}$. According to Eq.~(\ref{eq:dgsg}),
this implies that the $K \to 3\pi$ asymmetry 
is of the same order, as found within the SM.
It is interesting to note that values of $\Im \delta_\pm \sim 10^{-5}$ 
are consistent with the approximate flavor-symmetry scenario of Ref.~\cite{MM},
where $|\delta_\pm| \lsim \sin\theta_c m_s/m_{\tilde q}$.
In this framework one could therefore find a ``natural'' supersymmetric 
explanation for both $\varepsilon$ and $\varepsilon'$.

\begin{figure}[p]
\begin{center}
\leavevmode
\epsfxsize=12cm\epsfbox{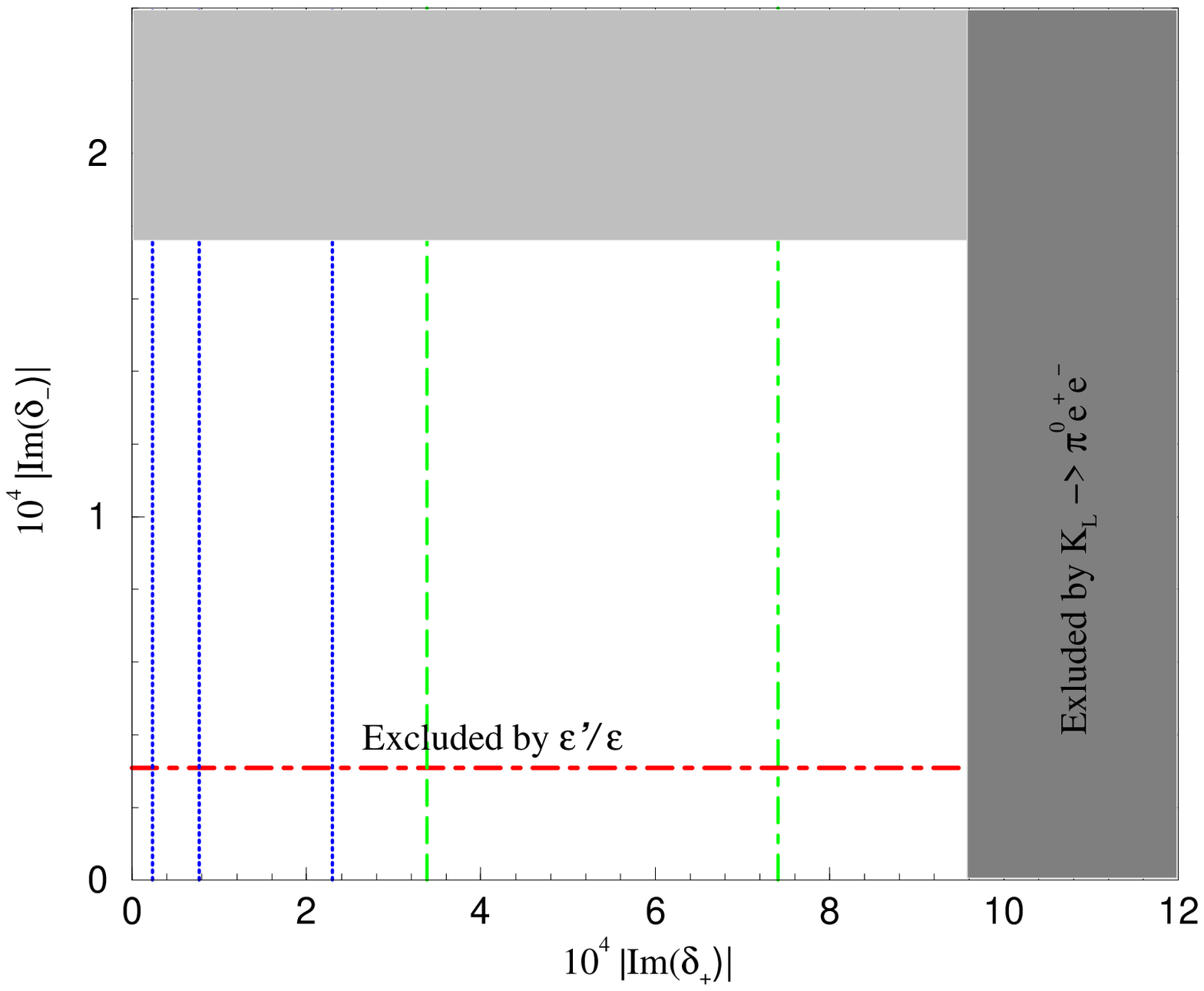}
\end{center}
\caption{\label{fig:unica} Constraints in the 
$\Im \delta_+$--$\Im \delta_-$ plane imposed by 
$\epse$, $\ep$ and ${\rm BR}(K_L \to \pi^0 e^+ e^-)$.
All bounds are obtained for $x_{gq}=1$ 
and scale linearly with $(500{\rm GeV}/m_{\tilde g})$. 
The vertical dotted lines correspond to  the 
$\ep$ constraint on ${\cal A}_{\rm 1-mag}$ for 
$\kappa_1=1.0$, $0.3$ and $0.1$ (from left to right); the other vertical lines
are obtained from the $\ep$ bound on
$({\cal A}_{\rm boxes}+{\cal A}_{\rm 2-mag})$
for $\delta_-=0$, $\Re \delta_+ = \Im \delta_+$ and
$B_+=\kappa_2=1$ (dashed) or $B_+=-\kappa_2=1$ (dot-dashed). 
The constraint from ${\rm BR}(K_L \to \pi^0 e^+ e^-)$
is obtained for $B_T=1$. 
The limits from $\epse$ are obtained for $B_{g2}=1$
and $(\epse)_{\rm SM}=0$ (horizontal dash-dotted line)
or the extreme case 
$(\epse)_{\rm SM} \leq - 120 \times 10^{-4}$ \protect\cite{newsoni}
(horizontal shadowed region). }
\end{figure}

To obtain larger values of the charge  asymmetry in $K \to 3\pi$ decays 
one has to relax the bound on   $\Im \delta_+$. To this purpose 
we note that $\epse$ put an explicit constraint only on 
 $\Im \delta_-$ but not on  $\Im \delta_+$, 
whereas ${\rm BR}(K_L \to \pi^0 e^+ e^-)$ and ${\cal A}_{\rm boxes}$ 
put upper bounds on $\Im \delta_+$ only  
at the level of $10^{-4}$. The strongest  limit
is set by the contribution of ${\cal A}_{\rm 1-mag}$ to $\ep$, since this term
is linear in $\Im \delta_+$ (the quadratic terms may become competitive only
for $\Re \delta/\Im \delta \gg 1$ or if $\kappa_1\ll\kappa_2$).
This contribution is subject to a large uncertainty which is parametrized by
$\kappa_1$. Therefore one can relax the bound by taking for $\kappa_1$ a value
sensibly smaller than one, as for example done in Ref.~\cite{mura2}.
In Fig.~\ref{fig:unica}, 
we display the bounds obtained for $\kappa_1=1.0$, $0.3$ and $0.1$. 
Only in the latter (optimistic) case one may obtain $K\to 3 \pi$ 
asymmetries in the $10^{-4}$  range. 
This is very similar to what has been found 
in Ref.~\cite{mura2} for the CP asymmetries of hyperon decays. 
\par
Even accepting values of $\Im \delta_+ \sim 10^{-4}$,
it remains to be explained the large cancellation 
between $\Im (\delta_{LR}^D)_{21}$ and  $\Im (\delta_{RL}^D)_{21}$  
necessary to satisfy the $10^{-5}$ bound
on $\Im \delta_-$ imposed by $\epse$. 
An underlying mechanism forcing this cancellation exists, however,
in the $U(2)$ models considered in Ref.~\cite{BCS}.
\par
In principle one could relax the $\epse$ constraint 
on $\Im \delta_-$ by allowing other contributions to $\epse$
to be large. For example if one accept the striking
result of Ref.~\cite{newsoni},\footnote{This value 
completely disagrees in sign and size with experimental
measurements and with many theoretical estimates. We consider indeed 
premature to use it  in phenomenological
analyses.} $(\epse)_{\rm SM} \sim - 120 \times
10^{-4}$,
one  would need a large supersymmetric contribution, corresponding 
to $\Im \delta_-\sim 10^{-4}$ to reconcile the theory with the 
experimental number.
This, however, leads to a new fine tuning problem, 
because then the natural order of magnitude of $\epse$ would 
be $10^{-2}$ rather than $10^{-3}$.

\paragraph{5.} 
Large CP violating effects triggered by
misalignments between  quark and squark mass matrices are 
among the most promising phenomena to uncover Supersymmetry at 
low energy. Among the possible effects those driven by  
the chromomagnetic operator are particularly interesting since
they could  completely account for the already measured CP violating parameters,
$\ep$ and $\epse$. In this paper we have studied the possibility that effects 
of the  CMO are detectable from the  enhanced  asymmetry  
in $K \to 3\pi$ decays.  We find  that this is possible only
if  several conditions, on which we have 
a poor theoretical control, conspire in the same direction.
Moreover, even if this were the case, one should then face 
a fine-tuning  problem in $\epse$.  
 
Our analysis of supersymmetric CP-violating effects in 
 $K \to 3\pi$ decays is parallel to those recently performed
in $K\to \pi\pi\gamma$ \cite{CIP} and hyperon decays \cite{mura2}. 
In all these cases the conclusions are hampered by poor knowledge of some 
hadronic parameters, which in the future will hopefully be computed
on the lattice. We stress that this problem is absent 
(or at least much simpler)  in rare $K$ decays like 
$K_L\to\pi^0\nu\bar{\nu}(e^+e^-)$ \cite{BCIRS}, 
whose experimental investigation will definitely provide useful and 
unambiguous information about the nature of CP violation. 

\bigskip
We thank I. Mannelli for interesting 
discussions that stimulated us to start this work. 
G.I. thanks  LAL and the Laboratoire de Physique Th\'eorique 
of the CNRS at Universit\'e de Paris XI
for hospitality during the completion of this work.

\end{document}